%% file: FH214.tex
\documentclass[preprint,12pt]{elsarticle}

\include{paperdef}

\begin{document}

\begin{frontmatter}

\title{Precision calculations in the MSSM \\ Higgs-boson sector with \FH 2.14}

\author[a,b]{H.~Bahl\corref{author}}
\author[b]{T.~Hahn}
\author[c]{S.~Heinemeyer}
\author[b]{W.~Hollik}
\author[d]{\\S.~Paßehr}
\author[e]{H.~Rzehak}
\author[a]{G.~Weiglein}

\cortext[author]{Corresponding author.\\
\textit{E-mail address:} henning.bahl@desy.de}
\address[a]{DESY, Notkestra\ss{}e 85, D--22607 Hamburg, Germany}
\address[b]{Max-Planck-Institut f{\"u}r Physik, F\"ohringer Ring 6, D--80805
Munich, Germany}
\address[c]{Instituto de F\'isica Te\'orica, Universidad Aut\'onoma de Madrid
Cantoblanco, 28049 Madrid, Spain;\\
Campus of International Excellence UAM+CSIC, Cantoblanco, 28049, Madrid,
Spain;\\
Instituto de F\'isica de Cantabria (CSIC-UC), E--39005 Santander, Spain}
\address[d]{Sorbonne Universit\'e, CNRS, Laboratoire de Physique Th\'eorique et
Hautes \'Energies (LPTHE),
4~Place Jussieu, F--75252 Paris CEDEX 05, France}
\address[e]{CP3-Origins, University of Southern Denmark, Odense, Denmark}

\begin{abstract}

\input{00_abstract.tex}

\end{abstract}


\end{frontmatter}

\begin{textblock*}{13em}(\textwidth,1.5cm)
\raggedleft\noindent\footnotesize
CP3-Origins-2018-035 DNRF90\\
DESY~18-179 \\
IFT-UAM/CSIC-18-095 \\
MPP--2018--229
\end{textblock*}

\clearpage
\newgeometry{top=2.5cm,left=3cm,right=3cm,bottom=4cm,footskip=3em}

\section*{New version program summary}

\noindent
{\em Program Title:}
\FH
\\[0.5em]
{\em Licensing provisions:}
GPLv3
\\[0.5em]
{\em Programming language:}
Fortran, C, Mathematica
\\[0.5em]
{\em Journal reference of previous version:}
Comput. Phys. Comm. 180 (2009) 1426
\\[0.5em]
{\em Does the new version supersede the previous version?}
Yes.
\\[0.5em]
{\em Reasons for the new version:}
Improved calculations and code structure.
\\[0.5em]
{\em Summary of revisions:}
Apart from improvements discussed in other publications: implementation of
optional \DR\ renormalization of stop sector, adapted two-loop Higgs sector
renormalization, implementation of full non-degenerate threshold corrections,
interpolation of EFT calculation for complex parameters, better code structure.
\\[0.5em]
{\em Nature of problem:}
The Minimal Supersymmetric Standard Model (MSSM) allows predictions for
the masses and mixings of the Higgs bosons in terms of a few relevant
parameters.  Therefore, comparisons to experimental data provide
constraints on the parameter space.  To fully profit from the experimental
precision, a comparable level of precision is needed for the theoretical
prediction.
\\[0.5em]
{\em Solution method:}
State-of-the-art fixed-order and effective-field-theory calculations are
combined to obtain a precise prediction for small as well as large
supersymmetry scales.

\clearpage
\tableofcontents



\section{Introduction}
\label{sec:01_intro}

\input{01_intro.tex}


\section{\FH overview}
\label{sec:02_FHintro}

\input{02_FHintro.tex}


\section{Improvements of the fixed-order calculation}
\label{sec:03_fixedorder}

\input{03_fixedorder.tex}


\section{Improvements of the EFT calculation}
\label{sec:04_EFT}

\input{04_EFT.tex}


\section{Improvements of code structure}
\label{sec:05_code}

\input{05_code.tex}


\section{Numerical results}
\label{sec:06_results}

\input{06_results.tex}


\section{Conclusions}
\label{sec:07_conclusion}

\input{07_conclusion.tex}


\section*{Acknowledgments}
\sloppy{
We thank E.~Bagnaschi, P.~Slavich, and I.~Sobolev for useful discussions and
relentless testing.
We thank E.~Bagnaschi, P.~Slavich, D.~Stöckinger, K.~Williams, and L.~Zeune for
contributions to the code.
The work of S.H.\ is supported in part by the MEINCOP Spain under
contract FPA2016-78022-P, in part by the ``Spanish Agencia Estatal de
Investigaci\'on'' (AEI) and the EU ``Fondo Europeo de Desarrollo
Regional'' (FEDER) through the project FPA2016-78022-P, in part by the
``Spanish Red Consolider MultiDark'' FPA2017-90566-REDC, and in part by
the AEI through the grant IFT Centro de Excelencia Severo Ochoa
SEV-2016-0597.
G.W.\ acknowledges support by the DFG through the  SFB 676 ``Particles, Strings
and the Early Universe''.
The work of S.P.\ is supported by the ANR grant ``HiggsAutomator'' (ANR-15-CE31-0002).
HR's work is partially funded by the Danish National Research Foundation, grant
number DNRF90.
The authors would like to express special thanks to the Mainz
Institute for Theoretical Physics (MITP) for its hospitality and support.
}


\appendix



\newpage

\begingroup
\setstretch{0}
\setlength{\bibsep}{3.0pt}
\renewcommand{\bibfont}{\footnotesize}
\section*{References}
\bibliographystyle{elsarticle-num}
\bibliography{biblio}{}
\endgroup

\end{document}

%% file: paperdef.tex
\usepackage[utf8x]{inputenc}      
\usepackage{amsmath,amssymb,mathtools}
\usepackage{alltt}
\usepackage[T1]{fontenc}          
\usepackage{booktabs,tabularx}
\usepackage{graphicx,subfig}
\usepackage{xspace}
\usepackage[usenames]{xcolor}\definecolor{fscolor}{RGB}{44,118,255}
\usepackage{soul}
\usepackage{geometry}
\usepackage{todonotes}
\usepackage{listings}
\usepackage[absolute]{textpos}
\usepackage[many]{tcolorbox}
\usepackage{xparse}
\usepackage[font=small,labelfont=bf,format=plain,margin=0.05\textwidth]{caption}
\usepackage{bbm}
\usepackage{tabularx}
\usepackage{setspace}
\usepackage{hyperref}
\usepackage[hyperpageref]{backref}
\makeatletter
\renewcommand\@makefntext[1]{\leftskip=.8em\hskip-.44em\@makefnmark#1}
\def\NAT@spacechar{\,}
\makeatother
\renewcommand*{\backref}[1]{}
\renewcommand*{\backrefalt}[4]{%
  \ifcase #1%
  \or [p\,#2]%
  \else [pp\,#2]%
  \fi%
}

\newif\ifbackrefshowonlyfirst
\backrefshowonlyfirsttrue
\makeatletter
\let\BR@direct@old@hyper@natlinkstart\hyper@natlinkstart
\renewcommand*{\hyper@natlinkstart}{\phantomsection\BR@direct@old@hyper@natlinkstart}
\let\BR@direct@oldBR@citex\BR@citex
\renewcommand*{\BR@citex}{\phantomsection\BR@direct@oldBR@citex}%

\long\def\hyper@page@BR@direct@ref#1#2#3{\hyperlink{#3}{#1}}

\ifx\backrefxxx\hyper@page@backref
    \let\backrefxxx\hyper@page@BR@direct@ref
    \ifbackrefshowonlyfirst
    \fi
\else
    \ifbackrefshowonlyfirst
    \fi
\fi

\patchcmd{\Hy@backout}{Doc-Start}{\@currentHref}{}{\errmessage{I can't seem to patch backref}}
\makeatother

\bibstyle{elsarticle-num}
\biboptions{sort&compress}
\newcommand\UH{$\mathbf{U}$-matrix\xspace}
\newcommand\ZH{$\mathbf{Z}$-matrix\xspace}
\newcommand\eg{e.g.\ }
\newcommand\ie{i.e.\ }

\newcommand\tb{\tan\beta}

\newcommand\DR{{\ensuremath{\overline{\text{DR}}}}}

\newcommand\OS{{\text{OS}}}

\newcommand\SM{{\text{SM}}}
\newcommand\MSSM{{\text{MSSM}}}

\newcommand\SUSY{{\text{SUSY}}}

\newcommand\FH{FeynHiggs\xspace}
\newcommand{\Fig}[1]{Fig.~\ref{#1}}
\newcommand{\Sec}[1]{Section~\ref{#1}}

\newcommand{\Eq}[1]{Eq.~(\ref{#1})}
\newcommand{\Eqs}[2]{Eqs.~(\ref{#1}) and (\ref{#2})}

\renewcommand{\Re}{\mathop{\mathrm{Re}}}
\renewcommand{\Im}{\mathop{\mathrm{Im}}}
\newcommand\ri{\mathrm{i}}
\newcommand\Code[1]{\ensuremath{\texttt{#1}}}

\newcommand\MW{M_W}

\newcommand\mgl{M_{\tilde{g\,}\!}}

\newcommand\cp{\ensuremath{{\cal CP}}}

\newcommand\msusy{M_\SUSY}

\newcommand\tev{\,\, \mathrm{TeV}}
\newcommand\gev{\,\, \mathrm{GeV}}
\newcommand\mev{\,\, \mathrm{MeV}}

\newcommand\order[1]{\ensuremath{\mathcal{O}(#1)}}
\newcommand\al{\alpha}
\newcommand\als{\al_s}
\newcommand\alt{\al_t}
\newcommand\alb{\al_b}

\newcounter{bla}

\journal{Computer Physics Communications}

%% file: 00_abstract.tex
We present an overview of the status and recent developments of \FH
(current version: 2.14.3) since version 2.12.2.  The main purpose of \FH
is the calculation of the Higgs-boson masses and other physical
observables in the MSSM. For a precise prediction of the Higgs-boson
masses for low and high SUSY scales, state-of-the-art fixed-order and
effective-field-theory calculations are combined.  We first discuss
improvements of the fixed-order calculation, namely an optional \DR\
renormalization of the stop sector and a renormalization of the Higgs
sector ensuring the chosen input mass to be equivalent with the corresponding
physical mass.  Second, we describe improvements of the EFT
calculation, \ie an implementation of non-degenerate threshold
corrections as well as an interpolation for complex parameters.  Lastly,
we highlight some improvements of the code structure easing future
extensions of \FH to models beyond the MSSM.

%% file: 01_intro.tex
While the gauge sector of the Standard Model~(SM) is well
investigated, the experimental precision in the
Higgs
sector~\cite{Aad:2012tfa,Chatrchyan:2012xdj,Aad:2015zhl,Khachatryan:2016vau}
leaves significant room for physics beyond the~SM~(BSM).  One of the most
frequently discussed BSM theories containing an extended Higgs sector
is the Minimal Supersymmetric Standard
Model~(MSSM)~\cite{Fayet:1974pd,Fayet:1977yc,Nilles:1983ge,Haber:1984rc}
based upon the concept
of supersymmetry~(SUSY).  Apart from adding a superpartner to every
SM~degree of freedom, the~MSSM also introduces a second Higgs doublet
resulting in five physical Higgs bosons: at the tree level, these are
the~\cp-even~$h$ and~$H$~bosons, the~\cp-odd~$A$~boson as well as the
charged~$H^\pm$~bosons.  Owing to the underlying supersymmetry, the
Higgs sector is determined by only two additional non-SM
parameters at the tree level. Conventionally, they are chosen as
the ratio of the vacuum expectation values~(vevs) of the two
doublets,~$\tan\beta = v_2/v_1$, and the mass of the~$A$-boson,~$M_A$
(or the mass of the charged bosons,~$M_{H^\pm}$).

Consequently, the Higgs sector of the~MSSM is highly
predictive, \ie the increasingly precise measurements of the
properties of the Higgs boson discovered by the~\mbox{ATLAS}
and~CMS~collaborations~\cite{Aad:2012tfa,Chatrchyan:2012xdj} at the
Large Hadron Collider~(LHC) allow one to efficiently probe the
parameter space of the~MSSM by comparing the
high-precision measurements of Higgs-boson
properties to the corresponding theoretical predictions.

The most precisely measured quantity of the Higgs boson is its mass.
In order to perform a meaningful comparison between the measured value
and the theoretical prediction, it is essential to take
into account radiative corrections to the theory predictions,
since these have a large impact on the Higgs
sector of the~MSSM.  To obtain a prediction with an uncertainty
comparable to the experimental precision, much work has been dedicated
to the calculation of these corrections (for the~MSSM with real parameters
see~\cite{ Ellis:1990nz,Okada:1990vk,Okada:1990gg, Haber:1990aw,
Ellis:1991zd,Sasaki:1991qu,Chankowski:1991md,
Brignole:1992uf,Chankowski:1992er, Hempfling:1993qq,
Casas:1994us,Dabelstein:1994hb, Carena:1995bx,Carena:1995wu,
Pierce:1996zz,Haber:1996fp,
Heinemeyer:1998jw,Heinemeyer:1998kz,Zhang:1998bm,Heinemeyer:1998np,
Heinemeyer:1999be,Espinosa:1999zm, Carena:2000dp,Espinosa:2000df,
Espinosa:2001mm, Degrassi:2001yf,Martin:2001vx,Brignole:2001jy,
Brignole:2002bz,Martin:2002iu,Martin:2002wn,Degrassi:2002fi,
Dedes:2002dy, Dedes:2003km, Martin:2003qz,Martin:2003it,
Martin:2004kr,Allanach:2004rh,Heinemeyer:2004xw,Heinemeyer:2004gx,
Martin:2005qm,Martin:2005eg, Martin:2007pg, Harlander:2008ju,
Kant:2010tf, Hahn:2013ria,Draper:2013oza,
Borowka:2014wla,Bagnaschi:2014rsa,Degrassi:2014pfa,
Vega:2015fna,Borowka:2015ura,Lee:2015uza, Bahl:2016brp,Athron:2016fuq,
Bagnaschi:2017xid,Bahl:2017aev,Harlander:2017kuc, Athron:2017fvs,
Allanach:2018fif,Bahl:2018jom,Harlander:2018yhj}, for the MSSM with
complex parameters see~\cite{
Pilaftsis:1998pe,Demir:1999hj,Pilaftsis:1999qt,Choi:2000wz,
Carena:2000yi,Ibrahim:2000qj,Heinemeyer:2001qd,Ibrahim:2002zk,
Martin:2004kr,Frank:2006yh,Martin:2007pg,
Heinemeyer:2007aq,Hollik:2014wea,
Hollik:2014bua,Goodsell:2016udb,Passehr:2017ufr, Borowka:2018anu}).

Apart from tackling the actual calculations, there has also been a
major effort to make the results publicly available by providing
them in terms of easily usable computer
programs.

One such program is
\FH~\cite{Heinemeyer:1998yj,Heinemeyer:1998np,Hahn:2009zz,
Degrassi:2002fi,Frank:2006yh,Hahn:2013ria,Bahl:2016brp,Bahl:2017aev}, which is
available at \Code{http://feynhiggs.de}.
Its main purpose is the calculation of the Higgs-boson masses in
the~MSSM. In addition, it provides precise predictions for various other phenomenologically
relevant observables. It has become a standard tool that is used for instance by
the LHC~Higgs Cross-Section Working Group~\cite{deFlorian:2016spz,Heinemeyer:2013tqa}
for the calculation of the MSSM Higgs boson masses and branching ratios.
For the calculation of the Higgs boson masses a combined approach of
fixed-order Feynman-diagrammatic and effective-field-theory (EFT)
calculations is employed.  In this paper, we describe recent updates
of the code released in versions~2.13.0~through~2.14.3. These
mainly improve the calculation of the Higgs boson masses and self-energy corrections,
which also enter the calculation of other observables like \eg the Higgs branching ratios.
We will therefore focus on the evaluation of these corrections in the present work.\footnote{A detailed comparison with other codes that evaluate the Higgs boson mass spectrum~\cite{Allanach:2001kg,Porod:2003um,Athron:2014yba,Harlander:2017kuc} or that can use the spectrum calculated by FeynHiggs to obtain the Higgs decay widths or production cross-sections~\cite{Djouadi:1997yw,Harlander:2012pb,Liebler:2016ceh} is beyond the scope of this paper.}
The improvements affect the
diagrammatic calculation~(see \Sec{sec:03_fixedorder}) as well as
the~EFT calculation~(see \Sec{sec:04_EFT}). Moreover,
we review the observables calculated by \FH. We discuss
improvements of the code structure~(see \Sec{sec:05_code}) and give a
short introduction to \FH and how to use
it~(see \Sec{sec:02_FHintro}). In the recent version~\FH~2.14.0, also
the pole-mass-determination procedure was improved. This issue
has been discussed in a separate publication \cite{Bahl:2018ykj}.

%% file: 02_FHintro.tex
In this section we give an overview about the various observables
  evaluated by \FH, together with the relevant references. The main
  emphasis is put on the Higgs-boson mass and self-energy calculations,
  which were in the focus of the updates of the recent years.

\subsection{Higgs boson mass spectrum}
\label{subsec:MHiggsCalc}

One of the main purposes of \FH is to provide predictions for the Higgs-boson
masses in the~MSSM.  The most direct approach is to calculate
higher-order corrections to the propagators of the Higgs bosons
performing a fixed-order Feynman-diagrammatic calculation.  \FH was
originally developed around this approach: It incorporates
full one-loop
contributions~\cite{Chankowski:1992er,Dabelstein:1994hb,Pierce:1996zz}
as well as the leading two-loop contributions\footnote{The two-loop
  self-energy corrections are computed in the approximation of
  vanishing electroweak gauge couplings and vanishing external
  momentum (see
  however~\cite{Borowka:2014wla,Borowka:2015ura,Borowka:2018anu} for
  studies going beyond this approximation).}
of~\order{\alt\als,\alb\als,\alt^2,\alt\alb,\alb^2}~\cite{
  Heinemeyer:1998yj,Heinemeyer:1998np,Degrassi:2001yf,
  Brignole:2001jy,Brignole:2002bz,Degrassi:2002fi,Dedes:2003km,
  Heinemeyer:2004xw,Frank:2006yh,Heinemeyer:2007aq,Hahn:2009zz,
  Hollik:2014wea,Hollik:2014bua,Hollik:2015ema,Hahn:2015gaa} to the
Higgs two-point functions. For these corrections, a mixed OS/$\DR$
renormalization scheme is employed (see~\cite{Frank:2006yh} for more
details).
The bottom-type Yukawa couplings include a resummation of the
$\tb$-enhanced terms (the ``$\Delta_b$
corrections'')~\cite{Hempfling:1993kv,Hall:1993gn,Carena:1994bv,Carena:1999py,Guasch:2003cv}
as detailed in~\cite{Brignole:2002bz,Dedes:2003km,Heinemeyer:2004xw}
(see also~\cite{Noth:2008tw,Noth:2010jy,Mihaila:2010mp} for corresponding
next-to-leading~order~(NLO) contributions).
The diagrammatic calculation allows one to take into
account complex parameters fully at the one-loop
level~\cite{Frank:2006yh} and
at~\order{\alt\als,\alt^2}~\cite{Heinemeyer:2007aq,Hollik:2014wea,
  Hollik:2014bua,Hahn:2015gaa} at the two-loop level (the phase
dependences of the other two-loop corrections are interpolated).
Moreover, non-minimal flavour violation can be considered at the
one-loop
level~\cite{AranaCatania:2011ak,Heinemeyer:2004by,Gomez:2014uha}.

The diagrammatic calculation captures all contributions at a given
order. This result contains logarithms involving some~SUSY~mass
  divided by the mass of a SM particle. For relatively low~SUSY
  scales, these logarithms are small and the fixed-order calculation
  is therefore expected to be precise. For a large separation between
  the SUSY scale and the electroweak scale, however, these logarithms
  become large.
Thus, they can spoil the convergence of the perturbative expansion,
rendering the fixed-order calculation inaccurate.

Effective-field-theory (EFT) techniques provide a tool to resum these
large logarithmic contributions to all
orders~\cite{Giudice:2011cg,Draper:2013oza,Bagnaschi:2014rsa,
Lee:2015uza,Vega:2015fna,Bagnaschi:2017xid,Bahl:2018jom,
Harlander:2018yhj}. The main idea is to integrate out some or all
heavy~SUSY~particles at a high scale.  The effective couplings
are then evolved down to the electroweak scale at which the Higgs mass
(or masses) are calculated, effectively resumming all large logarithms
that emerged from the masses of the heavy~SUSY~particles.  A
state-of-the-art~EFT calculation is available in
\FH~\cite{Hahn:2013ria,Bahl:2016brp,Bahl:2017aev}: based upon the
results of \cite{Bagnaschi:2014rsa,Vega:2015fna,Bagnaschi:2017xid}, it
includes full resummation of leading and next-to-leading
logarithms~(NLL) as well as~\order{\als,\alt}~resummation of
next-to-next-to-leading logarithms~(NNLL). Moreover, it allows one to
take into account light electroweakinos and gluinos by implementing
the corresponding low-energy thresholds and RGEs.
This logarithmic accuracy
level ensures a high precision for high~SUSY~scales. However,
since no higher-dimensional operators are included in
the~EFT~calculation, terms suppressed by the~SUSY~scale are
missed (see the discussion
in~\cite{Bagnaschi:2017xid}).  Therefore, the EFT calculation can
become inaccurate for low~SUSY~scales.

\bigskip

In order to ensure a precise prediction for low, intermediary, and
high~SUSY~scales, the fixed-order approach and the~EFT~approach are
combined in
\FH~\cite{Hahn:2013ria,Bahl:2016brp,Bahl:2017aev,Bahl:2018jom}. This
is achieved by adding the resummed logarithms obtained in
the~EFT~approach to the self-energies obtained in the fixed-order
approach and removing the double-counted logarithms by subtraction
terms.

Finally, the renormalized self-energies, $\hat\Sigma$, supplemented by
the resummed logarithms are used to obtain the pole masses of the Higgs
bosons.  For the neutral Higgs bosons this means that one has to find
the poles of the propagator matrix, whose inverse is given by
\begin{align}
\label{eq:propmatrix}
&\hat\Gamma_{hHA}(p^2)= \nonumber\\
& \ri\left[
  p^2 \mathbf{1} -
  \begin{pmatrix}m_h^2 & 0 & 0\\
    0 & m_H^2 & 0\\
    0 & 0 & m_A^2
    \end{pmatrix}+
\begin{pmatrix}
\hat\Sigma_{hh}(p^2) + \Delta_{hh}^\text{logs} &
\hat\Sigma_{hH}(p^2)  + \Delta_{hH}^\text{logs}& \hat\Sigma_{hA}(p^2)\\
\hat\Sigma_{hH}(p^2)  + \Delta_{hH}^\text{logs}&
\hat\Sigma_{HH}(p^2) + \Delta_{HH}^\text{logs} & \hat\Sigma_{HA}(p^2)\\
\hat\Sigma_{hA}(p^2) & \hat\Sigma_{HA}(p^2) &
\hat\Sigma_{AA}(p^2)
\end{pmatrix}\right].
\end{align}
The mixing with the neutral Goldstone boson and the~$Z$~boson
yields subleading two-loop contributions to the mass
predictions and is therefore
neglected. The~$\Delta$-terms contain the resummed logarithms,
obtained in the~EFT~approach, as well as the corresponding subtraction
terms.\footnote{The resummation of large logarithms is so far
  restricted to the $hh$, $hH$ and $HH$ self-energies. In the
  case of the SM as low-energy EFT, the resummation of logarithms in the
  $hH$ and $HH$ self-energies is approximated by
  assuming that the bulk of the correction originates from
  the top/stop sector. The coupling of the $H$ boson to top
  quarks is suppressed by $\tan\beta$ in the limit of large $M_A$.
  Therefore, the corrections to the $hH$ and $HH$ self-energies
  are obtained by dividing the correction
  to the $hh$ self-energy by $\tan\beta$ and $\tan^2\beta$, respectively
  (see \cite{Hahn:2013ria,Bahl:2018jom} for more details).
  The region of high $\tan\beta$ and low $M_A$, where the
  accuracy of this approximation is questionable, is already
  tightly constrained by experimental searches for heavy
  Higgs bosons.}
If all input
parameters are real,~$\hat\Sigma_{hA}$ and~$\hat\Sigma_{HA}$ vanish,
and the~\mbox{$(3\times 3)$} mixing is reduced to a~\mbox{$(2\times
  2)$}~mixing.

The real parts of the complex poles yield the physical Higgs-boson
masses. The masses are conventionally labelled
as~$M_{h_i}$~($i=1,2,3$) in the case of~\mbox{$(3\times 3)$}~mixing,
and as~$M_h$, $M_H$ and~$M_A$ in the case of~\mbox{$(2\times
  2)$}~mixing.

In order to treat external Higgs bosons on-shell (\eg in decay
rates), the (non-unitary)~\ZH is
calculated.~\cite{Chankowski:1992er,Heinemeyer:2001iy,Frank:2006yh,
Fuchs:2016swt,Fuchs:2017wkq} (see also~\cite{Domingo:2017rhb} and
Sect.~$5.3$ of~\cite{Fuchs:2015jwa}). It relates the tree-level mass
eigenstates to the external physical states.  Also an approximated
form of the~\ZH is given in the output, the~\UH. It is by default
defined as the unitary matrix diagonalizing the inverse propagator
matrix, \Eq{eq:propmatrix}, in the approximation of vanishing
momentum~\cite{Heinemeyer:2000fa,Heinemeyer:2001iy} and is used to
obtain effective couplings.

\FH furthermore provides an estimate of the remaining theoretical
uncertainties from unknown higher-order corrections
for all Higgs boson masses, for the~\ZH, and for
the~\UH~\cite{Degrassi:2002fi}.


\subsection{Other observables}
\label{subsec:other_observables}

In the following we list further (pseudo-)observables that are
   evaluated by\FH.
\footnote{The references focus on the corrections actually
    implemented into the code, but do not reflect the full status of the
    field of the corresponding available higher-order corrections.
    Reviews of Higgs boson production and decay,
    electroweak precision observables, EDM constraints and flavour constraints
    in the MSSM can be found in \cite{Spira:2016ztx}, \cite{Heinemeyer:2004gx},
    \cite{Pospelov:2005pr} and \cite{Buchalla:2008jp}, respectively.}
~The calculated Higgs masses and the~\ZH are used as input for the
prediction of various other observables in the MSSM.

The implemented
decay widths are summarized in Tab.~\ref{tab:decays}.\footnote{Various
  refinements to some of these decays, discussed
  in~\cite{Domingo:2018uim}, will soon be implemented in \FH.}
The NLO QCD corrections to the decays to massless gauge bosons are
  implemented in the heavy (s)quark limit. For the decays into massive
  vector bosons, the phrase ``reweighting of SM results''  refers to
  rescaling the SM result with the relevant
  coupling of the considered MSSM Higgs boson.

Furthermore, approximations (for fast
evaluation)---making use of tabulated SM results---of the main
Higgs production cross-sections for given~LHC~energies and~PDF~sets
are part of \FH, see Tab.~\ref{tab:xs}.
In this Table, the phrase ``reweighting of SM results''
refers to taking the SM cross section
(for the given value of the Higgs boson mass) and rescale it with the relevant
coupling of the considered MSSM Higgs boson, see~\cite{Hahn:2006my}
for more details. Information about the ``c-factor'' of the
$gg$~production cross section can be found in~\cite{Hahn:2010te} (and
references therein). The ``k-factor'' method applies higher-order
k-factors to the squared amplitude, taken
from~\cite{Spira:1995mt,Catani:2011kr}. ``Reweighting of THDM results''
refers to the application of the $\Delta_b$ corrections to the bottom
Yukawa coupling in the Two-Higgs-Doublet-Model cross section, which is
given in type II as a function of $M_{H^\pm}$ and $\tb$. More details
about the various cross sections can be found in the references given in
Tab.~\ref{tab:xs}. Moreover, the output contains a list of effective Higgs-boson
couplings.

In order to test the parameter space we also evaluate several
(pseudo-)observables that are connected to the Higgs-boson sector only
via higher-order corrections. In Tab.~\ref{tab:EWPOs} we list the
included electroweak precision observables. The SUSY corrections to
$\Delta\rho$ include full one-loop corrections, two-loop SUSY-QCD
corrections from gluons and gluinos as well as leading two-loop
electroweak corrections.  The leading two-loop SUSY corrections to
$\Delta r$ (and thus to $\MW$) and $\sin\theta_W^\text{eff,lept}$ are
incorporated via the $\rho$-parameter.  $\MW$ is calculated from
$\Delta r$ including full one-loop corrections and the SUSY two-loop
corrections in terms of $\Delta\rho$. For the calculation of
$\sin\theta_W^\text{eff,lept}$ only the one-loop SUSY corrections
through $\Delta r$ and the two-loop SUSY corrections through
$\Delta\rho$ are taken into account. From the SM side, the predictions
for $\Delta r$, $M_W$ and $\sin\theta_W^\text{eff,lept}$ contain all
higher-order corrections currently known (for more details
see \cite{Zeune:2014qpa}).  ``Partial 2L'' in the $(g-2)_\mu$ and the
EDMs predictions refers to the leading two-loop corrections. Details
can be found in the given literature.

\medskip

The flavor observables are given in Tab.~\ref{tab:flavour}.  For many
observables the corresponding SM~predictions are given, in order to
facilitate the comparison between~MSSM and SM~predictions.  For the
flavour observables, the recommendation is to use the values given in
the output only to be added to the best available SM~predictions
(which are not provided by \FH), as in: $O_{\MSSM,\text{best}} =
O_{\SM,\text{best}} + (O_{\MSSM,\text{FH}} - O_{\SM,\text{FH}})$.
Correspondingly, the references refer to the SUSY contribution only.

We stress again, as already mentioned above,
that the references listed in the Tables are not meant to
provide a comprehensive literature list for the quoted observable.
We list here only references containing corrections that are
implemented into \FH.

\begin{table}\centering
\begin{tabular}{|c|c|c|}
\hline
decay width / branching ratio &
	precision level & references \\
\hline
$h_i\to\gamma\gamma, gg$ &
LO + NLO QCD & \cite{Spira:1995rr,Spira:1997dg,Aglietti:2006tp, Benbrik:2012rm} \\ 
$h_i\to\gamma Z$ &
	LO & -- \\
$h_i\to ZZ, W^\pm W^\mp$ &
	reweighting of SM result &
\cite{Bredenstein:2006ha,Bredenstein:2006rh} \\
$h_i\to\bar f f$ &
	NLO & \cite{Williams:2011bu} \\
$H^\pm\to f f'$ &
	LO + NLO QCD & \cite{Djouadi:1994gf,Djouadi:1995gv} \\
$h_i\to\widetilde\chi_i^0\widetilde\chi_j^0$ &
	LO & -- \\
$h_i\to\widetilde\chi_i^\pm\widetilde\chi_j^\mp$ &
	LO & -- \\
$H^\pm\to\widetilde\chi_i^0\widetilde\chi_j^\pm$ &
	LO & -- \\
$h_i\to h_j Z$ &
	LO & -- \\
$H^\pm\to h_j W^\pm$ &
	LO &  \cite{Djouadi:1995gv}\\
$h_i\to h_j h_k$ &
	NLO + log resum. & \cite{Williams:2007dc,Williams:2011bu} \\
$h_i\to\tilde f \tilde f'$ &
	LO & -- \\
$H^\pm\to\tilde f_u \tilde f_d'$ &
	LO & \cite{Djouadi:1995gv} \\
\hline
\end{tabular}
\caption{\label{tab:decays}Higgs decay widths/branching ratios
  computed by \FH.  For decays including (excluding) loop corrections
  the \ZH\ (\UH) is employed by default, which includes propagator-type
  corrections at the same level of accuracy as the mass predictions.}
\end{table}

\begin{table}\centering
\begin{tabular}{|c|c|c|}
\hline
production cross section &
	precision level & references \\
\hline
$\bar b b\to h_i + X$ &
	reweighting of SM results &
\cite{Heinemeyer:2013tqa,deFlorian:2016spz}\\
$\bar b b\to h_i + X$ (one tagged $b$) &
	reweighting of SM results &
\cite{Harlander:2003ai,Heinemeyer:2013tqa,deFlorian:2016spz}\\
$g g\to h_i + X$ (c-factor) &
	reweighting of SM results &
\cite{Heinemeyer:2013tqa,deFlorian:2016spz}\\
$g g\to h_i + X$ (k-factor) &
	reweighting of SM results &
\cite{Heinemeyer:2013tqa,deFlorian:2016spz}\\
$q q\to q q h + X$ &
	reweighting of SM results &
\cite{Heinemeyer:2013tqa,deFlorian:2016spz}\\
$q q, g g\to t \bar t h_i + X$ &
	reweighting of SM results &
\cite{Heinemeyer:2013tqa,deFlorian:2016spz}\\
$q q\to W h_i + X$ &
	reweighting of SM results &
\cite{Heinemeyer:2013tqa,deFlorian:2016spz}\\
$q q\to Z h_i + X$ &
	reweighting of SM results &
\cite{Heinemeyer:2013tqa,deFlorian:2016spz}\\
$p p\to\tilde t_1\tilde t_1 h$ &
	LO &
\cite{Djouadi:1997xx,Kraus:2007privatecom}\\
$g b\to t H^-$ &
	reweighting of THDM results &
\cite{Berger:2003sm,Dittmaier:2009np,Heinemeyer:2013tqa,Flechl:2014wfa,Degrande:2015vpa,deFlorian:2016spz}
\\
$t\to H^+ b$ &
	LO + NLO QCD & \cite{Carena:1999py,Korner:2002fx} \\
\hline
\end{tabular}
\caption{\label{tab:xs}Higgs production cross-sections computed by
  \FH.}
\end{table}

\begin{table}\centering
\begin{tabular}{|c|c|c|}
\hline
EWPO &
	precision level &
	references \\
\hline
$\Delta\rho$ &
1L + 2L SUSY-QCD/EW &
\cite{Heinemeyer:2006px,Heinemeyer:2004gx,Heinemeyer:2013dia} \\
$\Delta r$ &
	1L + 2L SUSY-QCD/EW (full SM) &
	\cite{Heinemeyer:2006px,Heinemeyer:2004gx,Heinemeyer:2013dia,Stal:2015zca} \\
$\MW$ &
	1L + 2L SUSY-QCD/EW (full SM) &
	\cite{Heinemeyer:2006px,Heinemeyer:2004gx,Heinemeyer:2013dia,Stal:2015zca} \\
$\sin\theta_W^\text{eff,lept}$ &
	1L + 2L SUSY-QCD/EW (full SM) &
	\cite{Heinemeyer:2006px,Heinemeyer:2004gx,Heinemeyer:2007bw}
\\
$(g - 2)_\mu$ &
	1L + partial 2L &
	\cite{Degrassi:1998es,Heinemeyer:2003dq} \\
EDM of Th, n, and Hg &
	1L + partial 2L &
	\cite{Ibrahim:1997gj,Demir:2003js,Chang:1998uc,Olive:2005ru} \\
\hline
\end{tabular}
\caption{\label{tab:EWPOs}Electroweak precision observables computed
  by \FH. The abbreviation ``full SM'' is used to indicate that
  all known SM~corrections are taken into account.}
\end{table}

\begin{table}\centering
\begin{tabular}{|c|c|c|}
\hline
flavour observable &
	precision level & references  \\
\hline
$B\to X_s\gamma$ &
	LO & \cite{Hahn:2005qi} \\
$\Delta M_s$ &
	LO + NLO QCD & \cite{Buras:2001ra} \\
$B_s\to\mu^+\mu^-$ &
	LO + NLO QCD & \cite{Bobeth:2001jm} \\
\hline
\end{tabular}
\caption{\label{tab:flavour}flavour observables computed by \FH. All
implemented corrections allow one to take non-minimal flavour violation
into account.}
\end{table}


\subsection{Using \FH}
\label{subsec:usingFH}

\FH is mostly written in Fortran but can also be called from C/C++ and
Mathematica, or accessed from a Web interface. In order to build \FH a
Fortran and C~compiler and, to build the \FH~executables for
Mathematica, a working Mathematica/MathLink installation are needed.
The code has been thoroughly tested with gfortran, ifort, and pgf90 in
several versions on several platforms.

After downloading the latest tar file from \Code{http://feynhiggs.de},
the configuration and installation follow these steps:
\begin{samepage}
\begin{alltt}
   tar xvfz FeynHiggs-2.14.\(x\).tar.gz
   cd FeynHiggs-2.14.\(x\)
   ./configure
   make
   make install
\end{alltt}
\end{samepage}
After building the code, \FH\ provides several ways to use it:
\begin{itemize}
\item The \FH\ Fortran library \Code{libFH.a} can be linked to Fortran
or C/C++ programs, where the latter include \Code{CFeynHiggs.h}.

\item The \FH\ executable \Code{FeynHiggs} allows one to run \FH\ from
  the command-line.

\item The MathLink executable \Code{MFeynHiggs} allows one to call
  \FH\ from within a Mathematica session.
\end{itemize}

The Web interface at \Code{http://feynhiggs.de/fhucc}
allows one to run \FH\ without downloading it.

\FH\ receives its input parameters from an input file in either the
SLHA~\cite{Skands:2003cj,Allanach:2008qq}
or its native format, or directly through the API routines.  The contents
of the input file are read into a data structure called the \FH\ Record,
which can be thought of as an SLHA superstructure that also encodes loops
over parameters.  Routines to read an input file into the Record and step
through the loops in the Record are also available through the API.
The SLHA carries mostly \DR\ mass parameters, whereas FeynHiggs uses
mostly OS masses internally.  Care has to be taken if FeynHiggs is used
as the starting point of an SLHA chain as FeynHiggs presently cannot
convert its mass parameters to \DR\ before writing them to the SLHA;
this is indicated by a \DR\ scale of~0 in the corresponding block.

For more details, we refer to the manual pages which are included in the
tar file or are available at \Code{http://feynhiggs.de}.

%% file: 03_fixedorder.tex
In this section we describe improvements of the fixed-order calculation
starting from \FH 2.13 (released in early 2017).  The first
improvement is the implementation of an optional \DR\ renormalization of
the stop sector.  Second, we discuss an adaptation of the
renormalization in the Higgs sector at the two-loop level.


\subsection{Optional \texorpdfstring{$\,\overline{\!\text{DR}}$}{DRbar}
renormalization}
\label{subsec:DRbar}

\FH by default employs a mixed OS/\DR~renormalization scheme
(see~\cite{Frank:2006yh} for more details).  In particular, the
parameters of the stop/top sector are defined using OS~renormalization
conditions~\cite{Heinemeyer:2007aq} (stop masses and stop mixing
parameter~$X_t$).\footnote{The counterterm of~$X_t$ is fixed by
imposing a condition on the off-diagonal stop mass counter\-term~$\delta
m_{\tilde t_1 \tilde t_2}$ employing on-shell external momenta.
See \eg \cite{Heinemeyer:2007aq} for more details.}
\FH also offers the possibility to use \DR~input parameters, however.
Before the release of \FH~2.14, these were converted to OS~parameters
at the one-loop level.  The obtained OS~parameters were then used as
input for the rest of the calculation.  This procedure has the
advantage that a \DR~result and the default OS/\DR~result of \FH can
easily be compared.  If the calculation is performed identically
except for the renormalization schemes, the difference between the two
results can be interpreted as a part of the theoretical uncertainty.

As shown in~\cite{Bahl:2017aev}, this procedure is, however,
problematic if the fixed-order result is supplemented by a resummation
of large logarithms obtained in an EFT~approach.  The parameter
conversion induces additional logarithmic higher-order terms which can
become large for large SUSY~scales and therefore spoil the
resummation.  To circumvent this issue, an
optional \DR~renormalization of the stop sector was employed
in~\cite{Bahl:2017aev}.  Here we describe the practical implementation
of this optional renormalization scheme.

\begin{figure}[t]
\vspace{2ex}
\centering
\includegraphics[width=.85\linewidth]{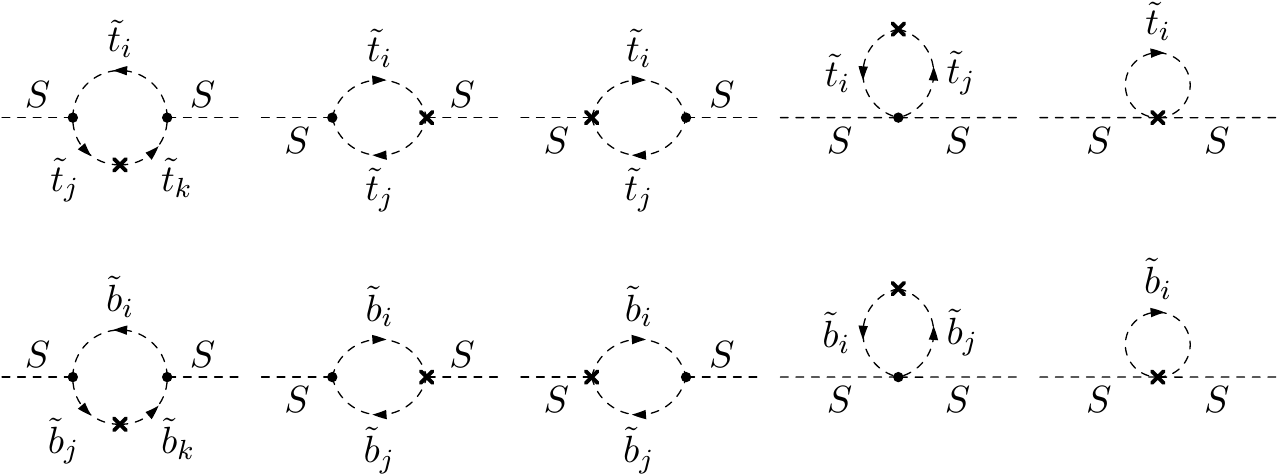}
\includegraphics[width=\linewidth]{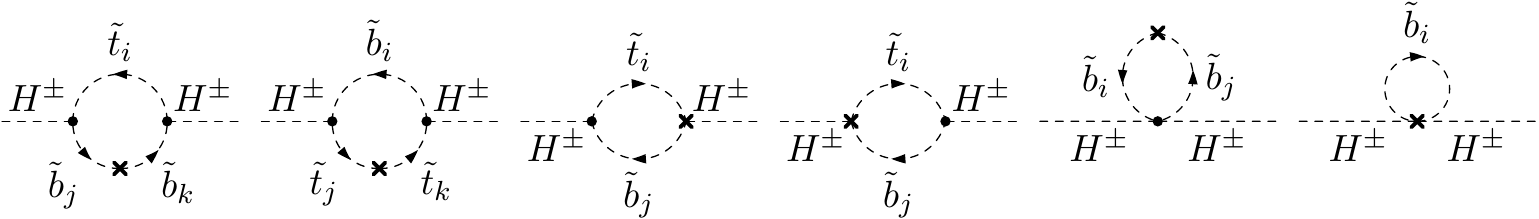}
\caption{\label{fig:TLsubloopren}Generic two-loop
subloop-renormalization diagrams appearing in the calculation of the
$\DR$~shifts (\mbox{$S=h,H,A$} and~\mbox{$i,j,k = 1,2$}).  Due to the
$SU(2)_L$~symmetry that relates the stop and sbottom sectors, also the
diagrams containing only bottom squarks yield contributions
involving stop counterterms.}
\vspace{2ex}
\end{figure}

This scheme is implemented with the stop-mass scale~\mbox{$M_S
= \sqrt{m_{\tilde t_1} m_{\tilde t_2}}$} as $\DR$~scale.  Inserting
the relation\footnote{The subscript ``fin'' indicates that only the
finite part of the OS~counterterm is taken into account.  The
UV-divergent part is cancelled by the corresponding \DR~counterterm.}
\begin{align}
X_t^\DR(M_S) = X_t^\OS + \delta^\OS X_t(M_S)\Big|_\text{fin}
\end{align}
and employing a Taylor expansion around $X_t^\OS$ we obtain
\begin{align}
\hat\Sigma(X_t^\DR(M_S)) = \hat\Sigma(X_t^\OS) +
  \left(\frac{\partial}{\partial X_t}\hat\Sigma\right)\cdot
  \delta^\OS X_t(M_S)\Big|_\text{fin},
\end{align}
where~$\hat\Sigma$ is a generic renormalized self-energy, \eg the
$hh$~self-energy.  The second term on the right-hand side corresponds
to the subloop-renormalization diagrams involving~$\delta X_t$ which
are depicted in \Fig{fig:TLsubloopren}.

In this way, changing the renormalization scheme and scale of the stop
sector becomes straightforward.\footnote{Another approach would have
been to replace the on-shell counterterms by \DR\ counterterms taking
into account the renormalization scale dependence.}
It amounts to the calculation of all
subloop-renormalization diagrams involving the stop mass or the stop
mixing counter\-terms with
the renormalization scale set equal to the stop mass scale $M_S$.
It should be noted that due to the $SU(2)_L$~gauge
symmetry also some sbottom counterterms depend on stop counterterms
(see \eg \cite{Heinemeyer:2004xw}).  Hence, also these
contributions have to be taken into account.
Adding the result to the existing self-energies with an OS~renormalized
stop sector, we have obtained the self-energies with a \DR-renormalized stop
sector.

\bigskip

This calculation is automated (see \Sec{sec:05_code}) and also works
for complex input parameters. In contrast,
the explicit conversion to OS~parameters had been implemented for real
input parameters only and was in practice applied to the absolute
value while the phase was left unchanged.
The new procedure is presently used in the stop sector of the mass
calculation; if the parameters of the sbottom sector are input in the
\DR~scheme, \FH still uses the explicit $\DR$/OS~conversion to obtain
the parameters renormalized in the mixed OS/\DR~scheme which is
employed for the sbottom sector~\cite{Brignole:2002bz}.  The explicit
conversion is likewise still used for the calculation of other
observables (\eg decay rates to scalar tops), with the exception of
the~\mbox{$h_i\to h_j h_k$} modes.  For the calculation of the latter,
the $\DR$~parameters of the stop sector are used in order to
consistently combine the
NLO~result~\cite{Williams:2007dc,Williams:2011bu} (see also
Tab.~\ref{tab:decays}) with a resummation of large logarithms. 


\bigskip

\subsection{Adapted renormalization of the Higgs sector}
\label{subsec:ImIm}

Another improvement concerns the renormalization of the Higgs sector
at the two-loop level.  If the mass of the \cp-odd Higgs boson~$A$ is
used as input mass (as done by default in the case of~\mbox{$(2\times
2$)}~mixing), the following OS~renormalization conditions are
employed,\footnote{At the two-loop level, all self-energy
contributions implemented in \FH are obtained by default in the limit
of vanishing external momentum.  Therefore, the counterterms are
adapted accordingly if they appear at the two-loop level
(see \eg \cite{Hollik:2014bua} for more details).  An exception are
the \order{\alt\als}~corrections for which optionally the full
momentum dependence can be taken into
account~\cite{Borowka:2014wla,Borowka:2015ura}.  Note that
the new additional contribution to the two-loop
counterterms~$\delta^{(2)} m_A^2$ and~$\delta^{(2)}m_{H^\pm}^2$,
discussed in this section, is not of~\order{\alt\als}.}\vspace{-.8ex}
\begin{align}
\delta^{(1)} m_A^2 ={}& \Re{\left[\Sigma^{(1)}_{AA}(m_A^2)\right]},
\label{eq:MAren1Lnew} \\
\delta^{(2)} m_A^2 ={}& \Re{\left[\Sigma^{(2)}_{AA}(m_A^2)\right]}
- \delta^{(1)}Z_{AA}\,\delta^{(1)}m_A^2 -
\delta^{(1)}Z_{GA}\,\delta^{(1)}m_{AG}^2\nonumber\\
&+
\Im{\left[\Sigma^{(1)\prime}_{AA}(m_A^2)\right]}
\Im{\left[\Sigma^{(1)}_{AA}(m_A^2)\right]},
\label{eq:MAren2Lnew}
\end{align}
where the $\delta^{(1)}Z$-s are one-loop field renormalization
constants (following the conventions of \cite{Hollik:2014bua}).

The physical mass squared, $M_A^2$, is given by the real part of the
corresponding propagator pole. In the absence of \cp-violation,
 \ie if all input parameters are real, this
pole is obtained by solving the equation
\begin{align}
p^2 - m_A^2 + \hat\Sigma_{AA}(p^2) = 0.
\end{align}
Expanding up to the two-loop level yields
\begin{align}
M_A^2 &= m_A^2 - \Re{\left[\hat\Sigma^{(1)}_{AA}(m_A^2)\right]} -
  \Re{\left[\hat\Sigma^{(2)}_{AA}(m_A^2)\right]} +
  \Re{\left[\hat\Sigma^{(1)\prime}_{AA}(m_A^2)
    \hat\Sigma^{(1)}_{AA}(m_A^2)\right]},
\end{align}
where the renormalized self-energies, marked by a hat, are given in
terms of the unrenormalized self-energies containing the
subloop renormalization and counterterms by
\begin{align}
\hat\Sigma^{(1)}_{AA}(m_A^2) &= \Sigma^{(1)}_{AA}(m_A^2) - \delta^{(1)}
m_A^2, \\
\hat\Sigma^{(2)}_{AA}(m_A^2) &= \Sigma^{(2)}_{AA}(m_A^2) -
\delta^{(1)}Z_{AA}\delta^{(1)}m_A^2 -
\delta^{(1)}Z_{AG}\delta^{(1)}m_{AG}^2 - \delta^{(2)}
m_A^2.
\end{align}
The superscript marks the loop order, and the prime is used to denote a
derivative with respect to~$p^2$.  Employing the conditions defined in
\Eqs{eq:MAren1Lnew}{eq:MAren2Lnew}, we straightforwardly obtain
\begin{equation}
M_A^2 = m_A^2,
\end{equation}
meaning that the input mass~$m_A$ is equivalent to the physical
mass~$M_A$. Before the release of \FH 2.14.0, the term in the
last line of \Eq{eq:MAren2Lnew} had been omitted.

If the charged Higgs boson mass~$m_{H^\pm}$ is used as input parameter
and renormalized on-shell (as done by default in the case
of~\mbox{$(3\times 3)$}~mixing in the neutral Higgs sector), its
two-loop counterterm is adapted accordingly,
\begin{align}
\delta^{(2)} m_{H^\pm}^2 ={}&
\Re{\left[\Sigma^{(2)}_{H^\pm H^\pm}(m_{H^\pm}^2)\right]}
- \delta^{(1)}Z_{H^\pm H^\pm}\,\delta^{(1)}m_{H^\pm}^2\nonumber\\
&-\frac{1}{2}\left(\delta^{(1)}Z_{G^\pm H^\pm}\,\delta^{(1)}m_{H^\pm G^\pm}+\delta^{(1)}Z_{G^\pm H^\pm}^{*}\,\delta^{(1)}m_{G^\pm H^\pm}\right)\nonumber\\
&+ \Im{\left[\Sigma^{(1)\prime}_{H^\pm H^\pm}(m_{H^\pm}^2)\right]}
   \Im{\left[\Sigma^{(1)}_{H^\pm H^\pm}(m_{H^\pm}^2)\right]},
   \label{eq:MHpren2Lnew}
\end{align}
whereas
\begin{align}
\delta^{(2)} m_A^2 = \delta^{(2)} m_{H^\pm}^2 - \delta^{(2)}\MW^2.
\end{align}
In the approximation of vanishing electroweak gauge couplings, as
employed for all two-loop corrections implemented in \FH, the two-loop
counterterm of the~$W$ boson mass~$\delta^{(2)}\MW^2$ is equal to
zero.

%% file: 04_EFT.tex
Apart from the fixed-order calculation, also the EFT~calculation that
is implemented in \FH has been improved.

\bigskip

The first advancement concerns the threshold corrections, which
appear at each of the matching scales (the calculation can contain up to four
different matching scales: the electroweak scale, the electroweakino scale, the
gluino scale and the sfermion scale).
Up to \FH~2.12.2, all threshold corrections were implemented in their
degenerate form. This means that at a threshold all particles which
are integrated out were assumed to have the same mass, which is
moreover equal to the matching scale.  As an example, in the EFT
calculation it was assumed that the soft SUSY-breaking masses of the
stop sector,~$M_{Q_3}$ and~$M_{U_3}$, are equal to each other. No such
assumptions have been made in the fixed-order calculation, however.
Therefore, the effect of non-degeneracy was captured in the fixed-order
calculation at the full one-loop level and the two-loop level in the limit
of vanishing electroweak gauge couplings.

In \FH~2.13.0, the full non-degenerate one-loop and the two-loop
threshold corrections of~\order{\alt\als}~\cite{Bagnaschi:2014rsa}
have been implemented. In \FH~2.14.1, also the
non-degenerate two-loop threshold corrections
of~\order{\alt^2}~\cite{Bagnaschi:2017xid} have been included.
This means that while \eg the two stops are still integrated out
at the same scale, their masses are not assumed to be equal anymore.
While this prescription yields precise results for $M_{Q_3} \sim M_{U_3}$,
more than one matching scale would be required in case of a large hierarchy
between the two soft SUSY-breaking masses

This facilitated to lift also a further restriction. Before, the
low-energy threshold of the gluino could only be taken into account in
the case of~LL and~NLL~resummation.  If NNLL~resummation was
activated, the gluino mass~$\mgl$ was set equal to the SUSY
scale~$\msusy$ in the EFT~calculation. The implementation of the
non-degenerate threshold correction of~\order{\alt\als} enables us to
set~$\mgl$ independently of~$\msusy$ in all relevant threshold
corrections.\footnote{Note that formally we would have to expand the
threshold corrections in~$\mgl/\msusy$ if the gluino remains in
the~EFT below the scale~$\msusy$. However, from a practical point of
view contributions of~\order{\mgl/\msusy} in the threshold corrections
are negligible in scenarios in which the gluino threshold has a
sizeable numerical impact (\ie if~\mbox{$\mgl\ll \msusy$}).  It should
be noted that logarithms involving the SUSY scale and the gluino mass
are not resummed for $\mgl>\msusy$.} No two-loop threshold corrections
need to be taken into account if the gluino is integrated out from the
effective theory below the scale~$\msusy$. This is due to the
couplings of the gluino: It couples either through a
quark--squark--gluino or a gluon--gluino--gluino vertex.  Therefore,
the gluino only contributes to the matching of the Higgs self-coupling
in the effective theory below the scale~$\msusy$---in which all
squarks are integrated out---at the three-loop level and beyond.
The~RGEs of the~EFT are modified already at the one-loop level,
however.  Corresponding one- and two-loop~RGEs are listed
in~\cite{Giudice:2011cg,Bahl:2016brp} and have been crosschecked using
SARAH, version 4~\cite{Staub:2013tta}. The modifications of the
three-loop~RGEs are unknown (except for the modification of the
three-loop running of the strong gauge
coupling~\cite{Clavelli:1996pz}).  The two-loop threshold corrections
are also valid for low mass gluinos.  Even so, the resummation of NNLL
logarithms has to be considered as an approximation in this case due
to the unknown three-loop~RGEs. Based on the finding that in the~SM
the effects from three-loop running are negligible
(see \eg \cite{Lee:2015uza}), it is, however, conceivable that also
the three-loop running in the~SM~plus~gluino is negligible.\footnote{
In addition, we checked that the implementation of the gluino
contribution to the three-loop beta function of the strong gauge
coupling~\cite{Clavelli:1996pz} leads to a negligible shift in $M_h$
of \order{1~\text{MeV}}.}  Taking into account the two-loop threshold
correction also valid for low mass gluinos, however, turns out to be
numerically relevant.  Therefore, using this two-loop threshold
correction together with the SM three-loop RGEs should be a good
estimate of the correct result.

\bigskip

As second improvement, an interpolation of the EFT~result was
introduced for complex parameters in \FH~2.13.0. A pure
EFT~calculation taking into account phases in the threshold corrections
has been performed in Ref.~\cite{Carena:2015uoe}, however, for the
hybrid approach no calculation for complex parameters is available at
the moment. Therefore, we follow the approach that is employed in \FH for those
fixed-order contributions which are only known for the case of real
parameters and interpolate the EFT~calculation in the case of complex
parameters. The interpolation is carried out for the Higgsino
mass parameter~$\mu$, the trilinear coupling in the stop sector~$A_t$,
and the gluino mass parameter~$M_3$, which are all allowed to take
complex values. The interpolation is performed by evaluating the
EFT~result at~$|P|$ and~$-|P|$ (\mbox{$P=\mu,A_t,M_3$}) and
afterwards linearly interpolating between the obtained values.

%% file: 05_code.tex
The code of \FH is structured in three parts:
\begin{itemize}
\item code hand-written for \FH

The `back bone' of \FH is of course written by hand.  Most code has
been developed specifically for \FH, with some adaptations from
external sources, \eg LoopTools~\cite{Hahn:1998yk} or
SLHALib~\cite{Hahn:2006nq}.  Code falling into this category includes
\begin{itemize}
\item all structural code: data structures, frontend, I/O, record
  handling, etc.
\item utility functions: matrix diagonalization, loop integrals,
  ordinary-differential-equation solver, etc.
\item contributions taken from the literature: the~EDMs, some of
  the~RGEs and threshold corrections in the EFT~sector, higher-order
  SM~parts of~$\Delta r$, etc., (for references
see Sect.~\ref{sec:02_FHintro}).
\end{itemize}

\item code generated from external expressions

\FH includes several contributions which originated from independent
projects and for which the original (typically large) expressions are
available, usually in Mathematica format: several of the two-loop
contributions to the Higgs self-energies, several ingredients of the
EFT~calculation, the muon~\mbox{$g - 2$}, the two-loop parts
of~$\Delta r$, etc.

This is already more practical than hand-coded expressions since
modifications (\eg a change in conventions) can be done in Mathematica
which is much easier and safer than search/replace in an editor.  Also
the code can be re-generated at any time and can be optimized, too.
On the other hand, it is nearly impossible to extend or significantly
change results implemented in this way.

\item code generated from calculations done in/for FeynHiggs

This mode is most convenient for perturbatively calculable quantities since
it allows full control over model content, particle selection,
resummations/$K$-factors, the renormalization prescription, etc.
Calculations done in this way can usually be generalized to other
models relatively straightforwardly.  Note that we do not pursue a
`generator generator'~approach as done in some other packages, \ie
even if our scripts ran (or were modified to run) with an `arbitrary'
model~file, the produced code would still need to be embedded in and
called from the main program, in which the inputs have to be
properly adjusted.

Calculations at this stage of automation can be found in the
`\Code{gen}'~subdirectory of \FH.

\newpage

Currently, it includes (for references see Sect.~\ref{sec:02_FHintro})
\begin{itemize}
\item the entire set of renormalized one-loop Higgs self-energies
  (\Code{gen/oneloop}),
\item the $\mathcal{O}(\alpha_t^2)$~contributions to the two-loop Higgs
  self-energies (\Code{gen/tlsp}),
\item the shifts at two-loop order from $\DR$~input parameters
  (\Code{gen/drbar}),
\item the shifts at two-loop order from finite $Z$~factors
  (\Code{gen/dzhfin}),
\item the one-loop decay rates
  (\Code{gen/decays}),
\item the one-loop corrections to~$\Delta_b$
  (\Code{gen/db}),
\item several flavour observables at the one-loop order
  (\Code{gen/bsg,bsll,dms}),
\item the one-loop MSSM contribution to~$\Delta\rho$
  (\Code{gen/deltarho})
\end{itemize}
The code-generation scripts generally follow the approach
of~\cite{Hahn:2015gaa} and, in case of improvements or bugfixes, can
be re-run with a few keystrokes.

(Another subdirectory, `\Code{gen/prod}', contains code for the
empirical fitting of cross-sections from tabulated data.  It falls
somewhat outside the sort of code generation described here and shall
not be discussed further.)

\end{itemize}

In the following we describe the main improvements in \FH
version~2.14.

The unrenormalized one-loop Higgs self-energies have been generated
with a high degree of automatization for all versions since~2.0.
Before \FH~2.14, however, the entire renormalization was hard-coded.
(At the time of the first implementation of the
self-energies, a model~file for the~MSSM
including the complete set of one-loop
counter-terms~\cite{Fritzsche:2013fta} did not yet exist.)
Old \FH~versions actually encoded various options of renormalization
schemes which were used for testing at that time.  The only
recommended scheme became the one used in the model file.  The
flags~\Code{fieldren} and~\Code{tanbren}, which selected these
schemes, were correspondingly dropped in~2.14.

The new procedure instead reads the renormalization (counter-terms
plus renormalization constants) from the model~file, making as few
assumptions as possible.  It needs to know the relevant flags
governing the renormalization, of course, such as~\Code{\$MHpInput},
which selects whether~$M_A$ or~$M_{H^+}$ is the input mass for the
Higgs sector, for which it generates the
necessary \texttt{if}~statements in the output.  Diagram computation
and code generation rely heavily on FeynArts~\cite{Hahn:2000kx} and
FormCalc~\cite{Hahn:1998yk}, and to achieve the level of automation we
desired, we had to enhance and add several of FormCalc's
code-generation functions \cite{Hahn:2019}.

New in~2.14 are also the two-loop shifts induced by the use of
$\DR$~input parameters and finite field-renormalization factors in the
one-loop Higgs self-energies.

Even though \FH does not (yet) go beyond the~MSSM in scope, there are
three~`models' used internally:~`\Code{mfv}' and~`\Code{nmfv}',
the~MSSM with minimal and non-minimal flavour-violation,
and~`\Code{gl}', the gaugeless version used \eg in the two-loop
calculations.  An important task was also to consolidate various
sources of Feynman~rules for the~MSSM which had grown over the years.

All one-loop self-energies are automatically split into the parts
corresponding to \FH' \Code{mssmpart} flag: $t/\tilde
t$; \mbox{$t/\tilde t + b/\tilde b$}; $f/\tilde f$; all, so that
individual sectors of the~MSSM can still be looked at even in the
presence of a generated renormalization.  Our code generation routines
are generic enough to deal with things such as different
renormalization schemes and simple extensions of the~MSSM but are also
to a certain extent model-aware,
\eg know how to simplify the \mbox{$(2\times 2)$} sfermion mixing
matrices, and are hence not directly applicable to `arbitrary' models.
Planned directions in this programme are the implementation of recent
two-loop results (\eg\cite{Passehr:2017ufr,Borowka:2018anu}) and
the extension to the~NMSSM based
on~\cite{Drechsel:2016jdg,Domingo:2017rhb,Domingo:2018uim}.

Finally, the adherence to the~FORTRAN~77 standard, kept mainly because
of~g77 (for many years the only free Fortran compiler), was dropped with
version~2.14.  Even though outwardly the code retains its fixed-format
`F77'~look, it uses many F90~idioms, in particular vector syntax.

While the numerical stability of the code is generally satisfactory,
some sections, for example the non-degenerate two-loop threshold
corrections of the EFT~results, can be affected by numerical
artefacts even in not-too-extreme corners of the parameter space. A
quadruple-precision version of FeynHiggs has been available
for some time (\Code{./configure {--}--quad}) but this naturally runs
vastly slower.  In~2.14.3 we reorganized many of the internal utility
functions, in particular the loop integrals, so that they compile
to either a double- or a quadruple-precision object depending on
the setting of a flag, and can now adjust higher precision for just
the neuralgic parts, which improves overall precision appreciably and
makes the slowdown hardly noticeable.  Quadruple precision
(\Code{REAL*16}, \Code{COMPLEX*32}) is currently available
with~gfortran and~ifort.  With~gfortran, the alternate
extended-precision type~\Code{REAL*10} can also be targeted, which is
realized in hardware on Intel~x86~chips, either overall
(\Code{./configure {--}--quad {--}--real10}) or just for the parts in
need of extra precision (\Code{./configure {--}--real10}).

%% file: 06_results.tex
In this Section, we present some exemplary results highlighting various
aspects of the improvements discussed above.  Other examples of the
improved Higgs-boson mass calculation are given in
\cite{Bahl:2017aev,Bahl:2018jom,Bahl:2018zmf,Bagnaschi:2018igf}.

\subsection{Improvements of the fixed-order calculation}
\label{subsec:res_FO}

First, we look at the improvements of the fixed-order calculation as
discussed in \Sec{sec:03_fixedorder}: the numerical impact of the
new optional \DR~renormalization on~$M_h$ obtained as a result of the
hybrid approach has already been presented
in~\cite{Bahl:2017aev}, and we do not repeat this discussion here.  We
will, however, investigate scenarios with complex \DR~input parameters
in \Sec{subsec:res_EFT}.

\begin{figure}
\centering
\begin{minipage}{.48\textwidth}\centering
\includegraphics[width=\textwidth]{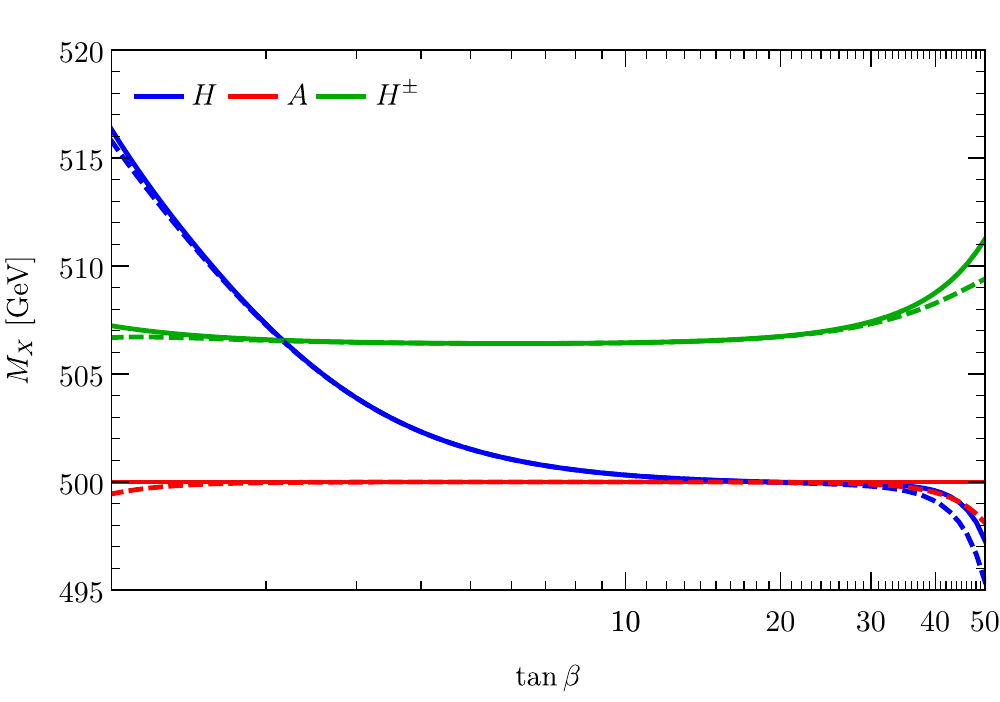}
\end{minipage}
\begin{minipage}{.48\textwidth}\centering
\includegraphics[width=\textwidth]{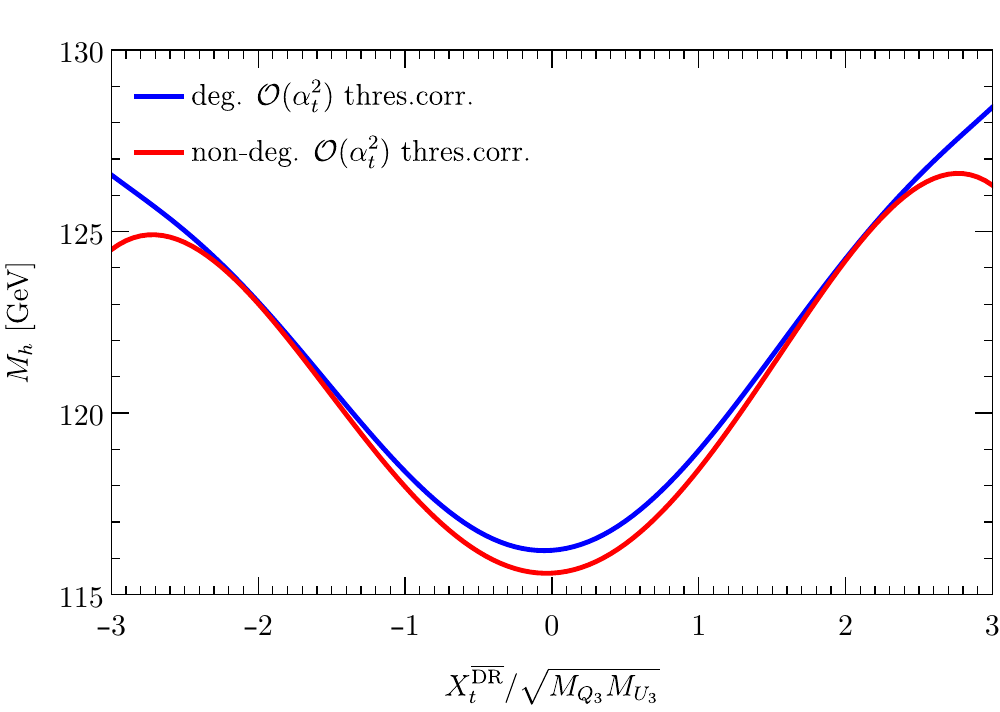}
\end{minipage}
\caption{Left: Masses of the non-SM-like Higgs bosons as a function
of~$\tan\beta$.  The results employing the adapted renormalization of
the Higgs sector~(solid) are compared to the results employing the old
renormalization~(dashed).  Right:~$M_h$ as a function
of~$X_t^\DR/\sqrt{M_{Q_3} M_{U_3}}$.  The results obtained using the
non-degenerate and the degenerate form of the threshold correction
of~\order{\alt^2} are compared.  (See text for the values of the
parameters.)}
\label{fig:imim_nondeg}
\end{figure}

The numerical effect of the adapted renormalization of the
Higgs sector, see \Sec{sec:03_fixedorder}, namely of the
additional term~$\Im[\Sigma^{(1)\prime}]\Im[\Sigma^{(1)}]$ in the
two-loop counterterm of the input mass in
Eq.~\eqref{eq:MAren2Lnew} or Eq.~\eqref{eq:MHpren2Lnew} is shown
in the left plot of \Fig{fig:imim_nondeg} for a scenario with
the input values~\mbox{$\msusy = 1\tev$} (common mass scale of squarks
and sleptons), \mbox{$X_t^\OS/\msusy=2$}, \mbox{$m_A = 500 \gev$},
and~\mbox{$\mu = - 500\gev$}.  The gaugino masses are
set to~\mbox{$M_1 = M_2 =500 \gev$}, and~\mbox{$ M_3 = 2.5\tev$}.  All
trilinear soft-breaking couplings apart from~$A_t$ are set to zero.

Due to the chosen mass pattern, the additional
term~$\Im[\Sigma^{(1)\prime}_{AA}(m_A^2)] \Im[\Sigma^{(1)}_{AA}(m_A^2)]$
only receives contributions from SM~particles.  One observes that the
term is negligible in the range~\mbox{$2\lesssim \tan\beta \lesssim
25$}.  For~\mbox{$\tan\beta \sim 1$}, where the coupling of the heavy
Higgs bosons to top~quarks is not suppressed, a small upward shift of
all three non-SM-like Higgs-boson masses is visible.  Similarly, one
finds a slightly larger upward shift for~\mbox{$\tan\beta\gtrsim 25$},
where the coupling of the heavy Higgs bosons to bottom~quarks becomes
large.  One also observes that with the adapted renormalization scheme
the physical mass of the~$A$-boson is, as expected, always equal to
the input mass~$m_A$.


\subsection{Improvements of the EFT calculation}
\label{subsec:res_EFT}

Next, we discuss the numerical impact of the improvements of the
EFT~calculation.  We first consider the effect of the
non-degenerate threshold corrections.  Since, as already
mentioned, the effect of non-degenerate particle masses was captured
exactly up to the level of two-loop corrections via the
fixed-order calculation before, the numerical impact of
those for scenarios with SUSY masses around the TeV scale is
quite small (\mbox{$\lesssim\order{100\mev}$}).

For multi-TeV SUSY masses larger effects can be observed, however.  As
an example, we investigate a scenario in which all soft-breaking
masses, the mass of the \cp-odd Higgs boson,~$m_A$, and the
Higgsino mass parameter~$\mu$ are set equal to~\mbox{$\msusy =
5 \tev$}.  Only the soft-breaking mass~$M_{U_3}$ in the stop sector is
chosen differently,~\mbox{$M_{U_3} = \msusy/4$}, to generate a large
non-degeneracy in the stop sector.  $\tan\beta$ is set equal to~$10$.
In the right plot of \Fig{fig:imim_nondeg}, we show~$M_h$ as a
function of~$X_t^\DR/\sqrt{M_{Q_3} M_{U_3}}$, comparing the results
obtained with the degenerate and the non-degenerate threshold
corrections of~\order{\alt^2}.  Due to the multi-TeV SUSY scale we
observe a downwards shift of~${\sim}\, 1 \gev$ for vanishing stop
mixing.  Moreover, we see that the values of~$X_t^\DR$
maximizing~$M_h$ are shifted away from the expected value
of~\mbox{$\lvert X_t^\DR/\sqrt{M_{Q_3} M_{U_3}}\rvert\sim\sqrt{6}$} if
the degenerate threshold correction of~\order{\alt^2} is used.  This
effect was especially relevant for the studies conducted
in~\cite{Bagnaschi:2018igf}.

For a further example showing the impact of the non-degenerate
threshold corrections of~\order{\alt^2}, we refer
to~\cite{Bahl:2018jom} where scenarios with low~$m_A$ are investigated
and shifts of up to~$6\gev$ in the prediction of $M_h$
have been found between results obtained
using the degenerate and non-degenerate threshold corrections
of~\order{\alt^2}.

\medskip

As second improvement we investigate the interpolation of the
EFT~result for the case of complex input parameters.  We compare
three methods to handle complex parameters in the EFT~calculation:
using the real part of the complex parameter as input,
using its absolute value as input, and the interpolation method
described in \Sec{sec:04_EFT}.  For the investigation, we use a
scenario like the one in the right plot of
\Fig{fig:imim_nondeg} but with~\mbox{$M_{U_3} = \msusy = 2 \tev$}.  In
addition, we allow for nonzero phases of~$A_t$ and~$M_3$.

\begin{figure}[t!]
\centering
\begin{minipage}{.48\textwidth}\centering
\includegraphics[width=\textwidth]{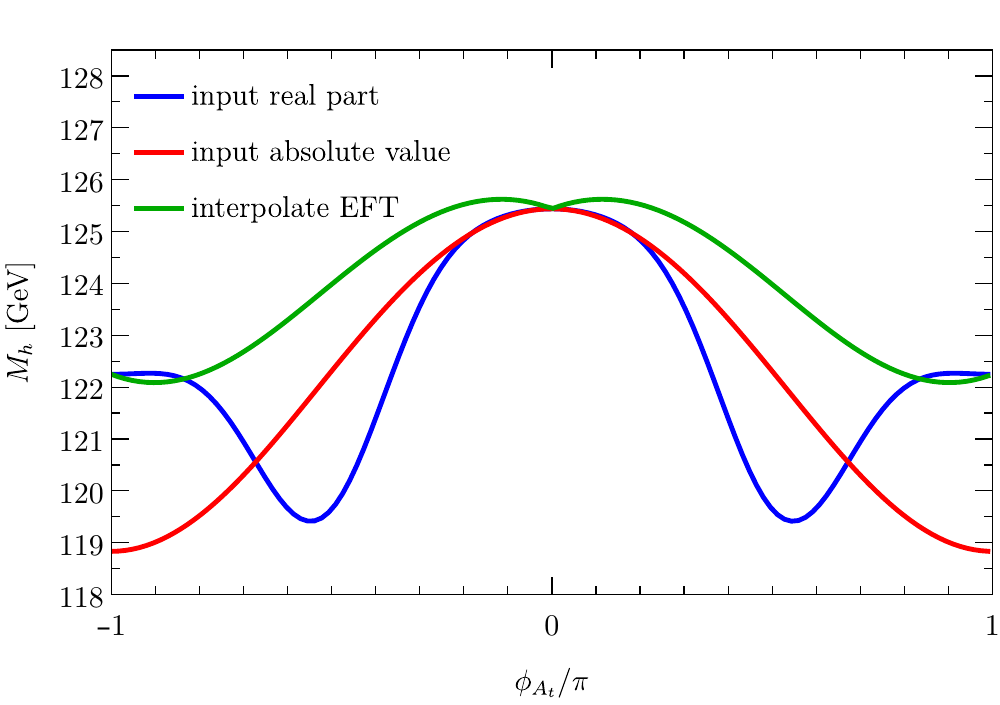}
\end{minipage}
\begin{minipage}{.48\textwidth}\centering
\includegraphics[width=\textwidth]{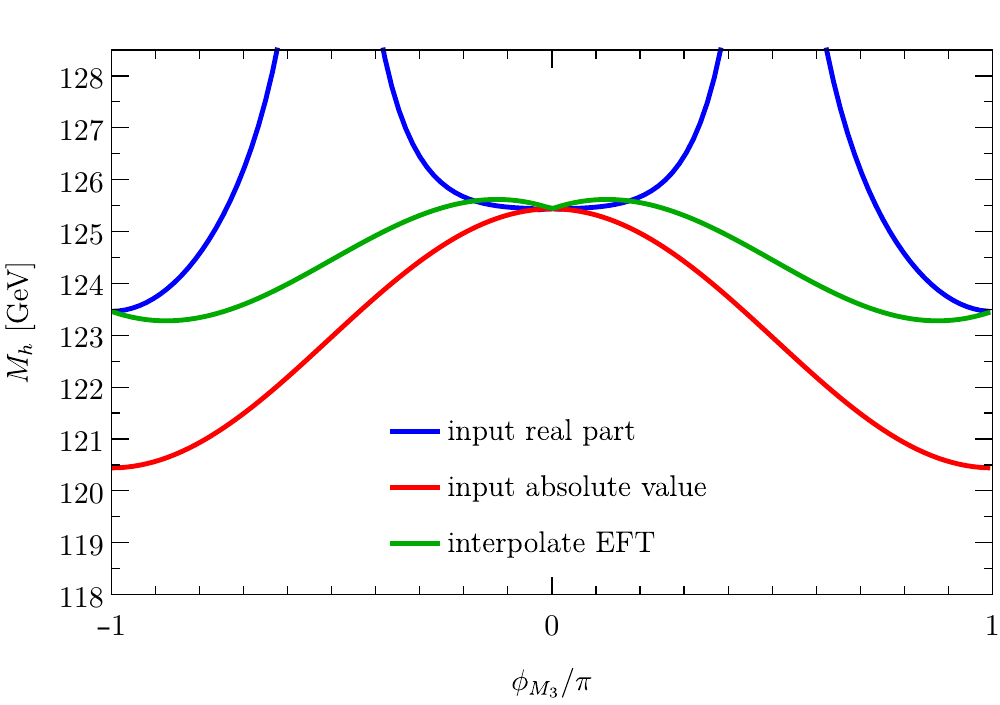}
\end{minipage}
\caption{\label{fig:interpolate}
Comparison of results with and without interpolation of the
EFT~result for complex parameters. The input parameters~\mbox{$\msusy
= 2\tev$}, \mbox{$\tan\beta = 10$} and~\mbox{$X_t^\DR/\msusy = \sqrt
6$} are chosen.  Left:~$M_h$ as a function of~$\phi_{A_t}$.
Right:~$M_h$ as a function of~$\phi_{M_3}$.}
\vspace{-1.1ex}
\end{figure}

In the left panel of \Fig{fig:interpolate} we vary the phase of~$A_t$
between~$-\pi$~and~$\pi$ and observe shifts in~$M_h$ of up to~$3\gev$
for~\mbox{$\phi_{A_t}\sim \pm\frac{\pi}{4}$}.  Cutting off the
imaginary part of~$A_t$ leads to values of~$M_h$ which are
similar to those obtained from the interpolation
in~$\phi_{A_t}$ only close to~\mbox{$\phi_{A_t} =
0; \pm\pi$} where the imaginary part of~$A_t$ is
small. For phases in between, the predicted values of~$M_h$ are
smaller compared to those obtained from the interpolation. Using the
absolute value conversely
works better for~\mbox{$\lvert\phi_{A_t}\rvert\lesssim 0.7$} but is
worse, as expected, for~\mbox{$\phi_{A_t} \sim\pm\pi$}.  Since the
one-loop threshold correction involves only even powers of~$X_t$, and
in the investigated scenario~$A_t$ is similar in size to~$X_t$ due to
the relatively high value of~$\tan\beta$, the dominant contribution
causing these shifts is the threshold correction of~\order{\alt\als}.

This is confirmed by the right plot of \Fig{fig:interpolate}, showing
a variation of the gluino phase~$\phi_{M_3}$.  The threshold
correction of~\order{\alt\als} is a function of~$X_t/M_3$.  Therefore,
a variation of~$\phi_{M_3}$ is comparable to a variation
of~$\phi_{A_t}$, as observable in the plots.  Cutting off the
imaginary part of~$M_3$ is not a good approximation here since~$M_3$
appears in the denominator and its real part approaches zero
for~\mbox{$\phi_{M_3}\sim\pm\frac{\pi}{4}$}.

The different treatment of the phases is formally of
NNLL order, since at the one- and two-loop level in
the fixed-order calculation the phase dependence
is taken into account without approximation.
The remarkably large size of the effect is compatible
with the shifts caused by overall NNLL resummation
found in \cite{Bahl:2016brp}. In contrast, a
variation of~$\phi_\mu$ leads only to very small shifts well
below~$1\gev$.

The results show that an interpolation of the EFT result yields
more reliable results than just using the real part or absolute value
of the complex parameters. Nevertheless, the displayed results
motivate an improved EFT calculation taking the phases fully into account.
We leave that for future work.

The plots shown in \Fig{fig:interpolate} are also examples of
scenarios with complex \DR~input parameters.  The conversion between
the \DR~input parameters and the internally used OS~parameters, as
employed in earlier \FH~versions, was in contrast not applicable to
the case of complex parameters (\ie the phases were not converted to
the OS~scheme).

%% file: 07_conclusion.tex
After presenting a short overview over the Fortran code \FH (available at
\Code{http://feynhiggs.de}), whose
main purpose is to provide precise numerical predictions for
 observables in the Higgs sector of
the~MSSM, we discussed various improvements in the calculation of the
Higgs spectrum.  For the prediction of the Higgs-boson masses, a
diagrammatic fixed-order calculation---accurate for low
SUSY~scales---is combined with an EFT~calculation---accurate for high
SUSY~scales---in order to provide an accurate result
also for intermediate scales.\footnote{The uncertainty in the
prediction of $M_h$ will be discussed in a future publication
which will compare the different approaches in detail.}

We first discussed improvements of the fixed-order calculation. We
explained the implementation of an
alternative \DR~renormalization of the stop sector
(allowing one to input also complex \DR~parameters).
Moreover, we showed how the two-loop renormalization of the Higgs
sector is adapted in order to ensure that the input Higgs mass is
equal to the corresponding physical mass.  Numerically, this change of
the renormalization scheme has been relevant in the considered scenario only for
very low or very high
values of~$\tan\beta$.

Then, we addressed the improvements of the EFT~calculation. We
described the implementation of threshold corrections valid for
arbitrary masses of the decoupled particles and showed that this can
lead to sizeable numerical effects as compared to the result using
degenerate threshold corrections \eg in the case of a large
separation between the two stop masses.  Moreover, we explained how
the EFT~calculation is interpolated in the case of complex input
parameters. The numerical effects
of the phase variations can be important if the imaginary parts
of the respective parameters are sufficiently large.

We furthermore highlighted several improvements of the code structure, which
are not directly visible for the user but should allow for an easier
development and extension of \FH in the future.